\documentclass[aps,prl,reprint,showpacs]{revtex4-1}

\usepackage{graphicx}
\usepackage{subfigure}
\usepackage{bbm}
\usepackage{amsmath}
\usepackage[pdftitle={Ion-photon quantum interface for state mapping and entanglement},pdfauthor={Christoph Kurz},pdfdisplaydoctitle=false,bookmarks=false,pdflang={en-US}]{hyperref}
\newcommand{\ket}[1]{\mathopen|#1\rangle}
\newcommand{\bra}[1]{\langle#1\mathclose|}
\newcommand{\Ca}{^{40}\text{Ca}^+}
\newcommand{\sstate}{\text{S}_{1/2}}
\newcommand{\Pstate}{\text{P}_{3/2}}
\newcommand{\Dstate}{\text{D}_{5/2}}
\newcommand{\half}{\tfrac{1}{2}}
\newcommand{\fivehalf}{\tfrac{5}{2}}

\begin{document}

\title{Programmable atom-photon quantum interface}

\author{Christoph Kurz}
\author{Pascal Eich}
\author{Michael Schug}
\author{Philipp M\"uller}
\author{J\"urgen Eschner}
\email{juergen.eschner@physik.uni-saarland.de}
\affiliation{Universit\"at des Saarlandes, Experimentalphysik, Campus E2 6, 66123 Saarbr\"ucken, Germany}

\begin{abstract}
We present the implementation of a programmable atom-photon quantum interface, employing a single trapped $^{40}$Ca$^+$ ion and single photons. Depending on its mode of operation, the interface serves as a bi-directional atom-photon quantum-state converter, as a source of entangled atom-photon states, or as a quantum frequency converter of single photons. The interface lends itself particularly to interfacing ions with SPDC-based single-photon or entangled-photon-pair sources. 
\end{abstract}


\maketitle

Quantum networks \cite{Kimble2008} integrate quantum-information processing and storage devices with quantum communication channels, allowing one to distribute quantum information between stationary network nodes through flying quantum bits (qubits). A platform that has proven suitable for this purpose is the combination of single trapped ions and single photons: while single ions allow for storing and processing quantum information \cite{Leibfried2003, Langer2005, Benhelm2008}, single photons are the only viable carriers of quantum information over long distance \cite{Weihs1998, Ma2012}. In order to interconvert stationary and flying qubits, quantum interfaces are needed. Different approaches for the experimental implementation of atom-photon quantum interfaces have emerged. These include objectives and lenses of high numerical aperture \cite{Tey2008, Vittorini2014}, deep parabolic mirrors \cite{Maiwald2009, Fischer2014}, and optical resonators \cite{Hood1998, Hennrich2000, Mundt2002, Keller2003}. With these systems, a number of essential quantum-networking building blocks have been realized, such as atom-photon entanglement \cite{Blinov2004, Volz2006}, atom-to-photon quantum-state mapping \cite{Boozer2007, Wilk2007, Stute2013}, heralded atom-atom entanglement \cite{Moehring2007, Hofmann2012}, quantum teleportation \cite{Olmschenk2009} and direct quantum-state transfer \cite{Noelleke2013}. We demonstrated further basic prerequisites of a quantum network such as the high-rate generation of single photons in a pure quantum state \cite{Kurz2013} and the direct photonic interaction between distant single ions \cite{Schug2013}. Recently, we implemented a protocol for high-fidelity heralded transfer of a photonic polarization qubit onto the qubit state of a single ion \cite{Kurz2014}. Here we report on a significant and comprehensive extension of that protocol, the realization of a programmable, bi-directional interface between a single ion and single photons, which depending on its mode of operation permits atom-to-photon or photon-to-atom quantum-state transfer, as well as generation of atom-photon entanglement; moreover, it is also suitable for quantum frequency conversion of single photons. Thereby the interface covers the essential operations required in quantum networks and especially for a quantum repeater \cite{Briegel1998}.


\subsection*{Interface principle}

\begin{figure}[bht]
   \includegraphics[width=0.84\columnwidth]{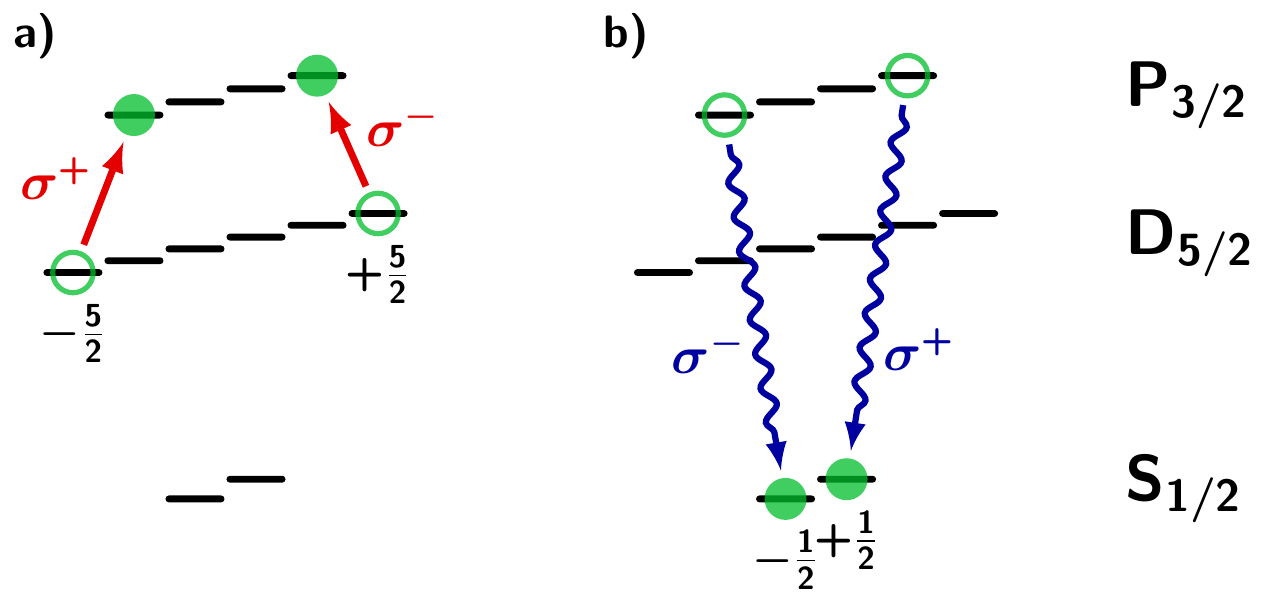}
   \caption{General experimental scheme. Black bars denote the Zeeman sublevels of the involved atomic states $\sstate$, $\Dstate$, and $\Pstate$. (a) Input state: the ion is coherently prepared in a superposition of the $\ket{\pm\fivehalf}$ Zeeman sublevels of $\Dstate$ (filled green circles), and a photon at 854\,nm in a superposition of the two circular polarizations is absorbed (red arrows). (b) Output state: a single 393\,nm photon in a superposition of the two circular polarizations is emitted (blue arrows), leaving the ion in the corresponding superposition of the $\ket{\pm\half}$ Zeeman sublevels of $\sstate$ (filled green circles).}
   \label{fig:scheme}
\end{figure}

Fig.\ \ref{fig:scheme} illustrates the general principle of the interface, experimentally implemented with the ground and excited states of a single $\Ca$ ion. We first create a coherent superposition of the two outer Zeeman sublevels of the metastable $\Dstate$ state,
\begin{equation}
   \ket{\psi_\text{D}}=\cos\frac{\vartheta_\text{D}}{2}\ket{\text{D},-\fivehalf} + \sin\frac{\vartheta_\text{D}}{2} e^{i\varphi_\text{D}}\ket{\text{D},+\fivehalf}~, \label{eqn:initialstateD}
\end{equation}
and expose the ion to resonant photons on the $\Dstate$ to $\Pstate$ transition at 854\,nm which are in a polarization state in the circular (R/L) basis,
\begin{equation}
   \ket{\psi_\text{854}}=\cos\frac{\vartheta_\text{854}}{2}\ket{\text{854,R}} + \sin\frac{\vartheta_\text{854}}{2} e^{i\varphi_{854}}\ket{\text{854,L}}~, \label{eqn:initialstate854}
\end{equation}
such that the input state of the interface is $\ket{\Psi_\text{in}} = \ket{\psi_\text{D}}\ket{\psi_\text{854}}$. Right-handed (R) and left-handed (L) circularly polarized 854\,nm photons may be absorbed on the $\sigma^+$ and $\sigma^-$ transition, respectively. Absorption of a photon triggers the emission of a single Raman photon on the $\Pstate$ to $\sstate$ transition at 393\,nm and transfers the ion to the $\sstate$ ground state. For detection along the quantization axis, the two possible transitions, $\sigma^+$ and $\sigma^-$, translate to R- and L-polarized 393\,nm photons, respectively. We hence obtain the ion-photon output state
\begin{equation}
   \ket{\Psi_\text{out}} = \cos\frac{\theta}{2}\ket{\text{S},-\half} \ket{\text{393,L}} + \sin\frac{\theta}{2} e^{i\phi}\ket{\text{S},+\half} \ket{\text{393,R}}~.  \label{eqn:entangledstate}
\end{equation}

The evolution of the system from $\ket{\Psi_\text{in}}$ into the entangled state $\ket{\Psi_\text{out}}$ is 
described by \cite{Schug2014}
\begin{eqnarray*}
	\ket{\Psi_\text{out}} = 
	\big( \ket{\text{S},-\half} \ket{\text{393,L}} && \bra{\text{D},-\fivehalf} \bra{\text{854,R}} +   \\ 
	+ \ket{\text{S},+\half} \ket{\text{393,R}} && \bra{\text{D},+\fivehalf} \bra{\text{854,L}} \big) \cdot 
	\ket{\Psi_\text{in}}
\end{eqnarray*}
and used to realize the different interface operations: (i) when $\ket{\psi_\text{D}}$ and $\ket{\psi_\text{854}}$ are fixed, entangled atom-photon pairs are created (entangler operation); (ii) when $\ket{\Psi_\text{out}}$ is projected by detecting the 393\,nm photon in a linear basis, the polarization state of the absorbed 854\,nm photon is mapped into an atomic qubit state in $\sstate$ (receiver operation); (iii) when the final atomic state is projected on a superposition of the Zeeman sublevels, the initial atomic state $\ket{\psi_\text{D}}$ is mapped into the polarization qubit of the output 393\,nm photon (sender operation); (iv) finally, when both atomic superpositions are fixed, any input photon will be converted into a frequency-converted output photon of the corresponding polarization (converter operation). Receiver operation has been shown in \cite{Kurz2014}, entangler and sender operation are demonstrated below. Quantum frequency conversion of photons from 854\,nm to 393\,nm is implicitly realized by these demonstrations.

\subsection*{Experimental procedure}

\begin{figure}[ht]
   \includegraphics[width=0.9\columnwidth]{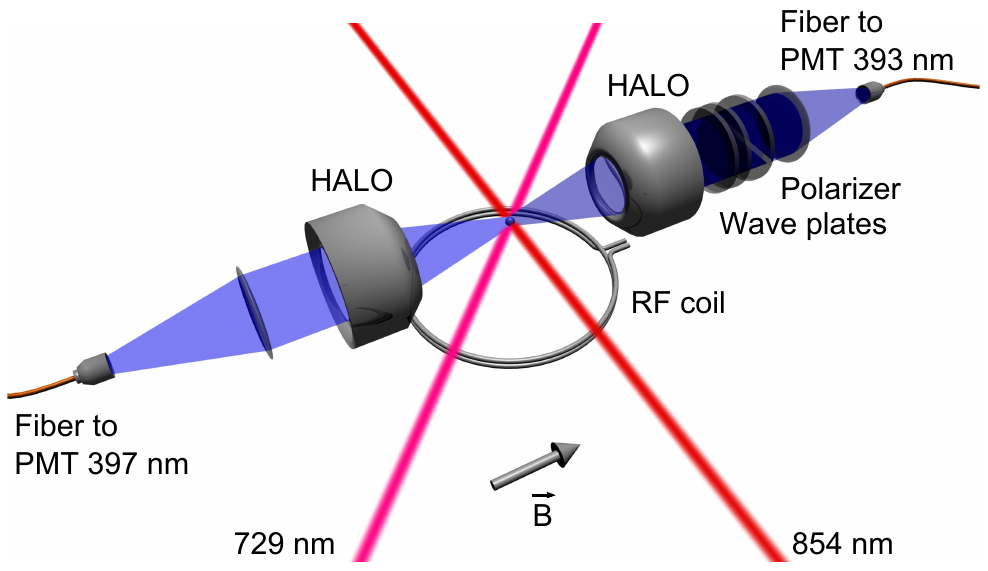}
   \caption{Experimental set-up. HALO: high-NA laser objective, PMT: photomultiplier tube, Wave plates: one half-wave and one quarter-wave plate, $\vec{B}$: magnetic-field direction. The ion is trapped between the HALOs.}
   \label{fig:setup}
\end{figure}

Our experimental set-up is sketched in Fig.\,\ref{fig:setup}. A single $\Ca$ ion is confined in a linear Paul trap and Doppler cooled by frequency-stabilized diode lasers \cite{Rohde2010}. A static magnetic field $B=2.8\,\text{G}$ defines the quantization axis. Atomic state preparation in $\ket{\psi_\text{D}}$ is performed in three steps \cite{Kurz2014, Schug2014}: first the ion is optically pumped to the $\ket{-\half}$ level of $\sstate$, then a radio-frequency (RF) $\pi/2$-pulse from a coil below the trap excites the magnetic-dipole transition between the $\sstate$ sublevels to form an equal superposition of $\ket{\pm\half}$, and finally two $\pi$-pulses from a narrow-band laser at 729\,nm coherently excite the $\ket{\pm\half}$ populations to the respective $\ket{\pm\fivehalf}$ levels in $\Dstate$. 

A laser at 854\,nm provides the polarized photons at 854\,nm. Since for ion-to-photon state transfer as well as for ion-photon entanglement (sender and entangler operation) the 854\,nm polarization is kept fixed, the laser propagation and polarization direction are both chosen orthogonal to the quantization axis. Photons at 393\,nm emitted by the ion are collected along the quantization axis through an in-vacuum high-NA laser objective (HALO, \cite{Gerber2009}). The photons are projected onto a chosen polarization and sent to photomultiplier tubes (PMTs) via multi-mode optical fibers. The output pulses of the PMTs, corresponding to the arrivals of single photons, are time-tagged for later processing. Another HALO collects fluorescence from cooling laser excitation. 

Analysis of the final $\sstate$ atomic state in the $\ket{\pm\half}$ basis is performed by shelving the $\ket{+\half}$ population in a sublevel of $\Dstate$ and illuminating the ion with the cooling light, which either reveals fluorescence (ion in $\ket{-\half}$) or not (ion in $\ket{+\half}$). For measuring in a superposition basis, an RF $\pi/2$-pulse with the respective phase effects a basis rotation before shelving. 

A typical experiment starts with the preparation of a given symmetric superposition in $\Dstate$ according to Eq.\,(\ref{eqn:initialstateD}), which has angles $\vartheta_\text{D}=\frac{\pi}{2}$, and $\varphi_\text{D}=\varphi_\text{729}+\omega_\text{L}t$. Here $\varphi_{729}$ is the (adjustable \cite{Schug2014}) phase between the two preparation pulses of the 729\,nm laser, and $\omega_\text{L}t$ is the phase that the superposition accumulates with time due to its Larmor precession \cite{Kurz2014, Schug2014}, $t=0$ marking its preparation. The ion is excited with 854\,nm photons of fixed linear polarization ($\vartheta_\text{854}=\frac{\pi}{2}$, $\varphi_{854}=\pi$), and detection events of 393\,nm photons in various polarization bases are recorded in the form of arrival-time histograms. For verifying ion-to-photon state transfer as well as ion-photon entanglement, the correlation of these photon detections with the projection of the atom onto a specific superposition state is analyzed. Detection of the 393\,nm photons with high time resolution ($\lesssim\,320$\,ps in our case) is crucial, in order to remove the frequency distinguishibility of the two Raman scattering paths \cite{Togan2010}. 

\begin{figure}[ht]
   \makeatletter\renewcommand{\@thesubfigure}{\hspace{11mm}\smash{\thesubfigure}}\makeatother
   \subfigure[]{\includegraphics[width=0.9\columnwidth]{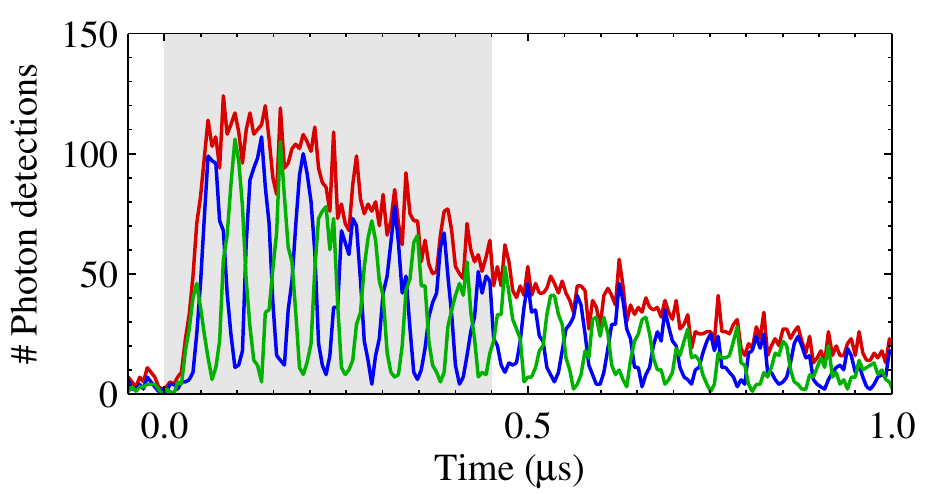} \label{fig:oscphotons}}
   \makeatletter\renewcommand{\@thesubfigure}{\hspace{11mm}\smash{\thesubfigure}}\makeatother
   \subfigure[]{\includegraphics[width=0.9\columnwidth]{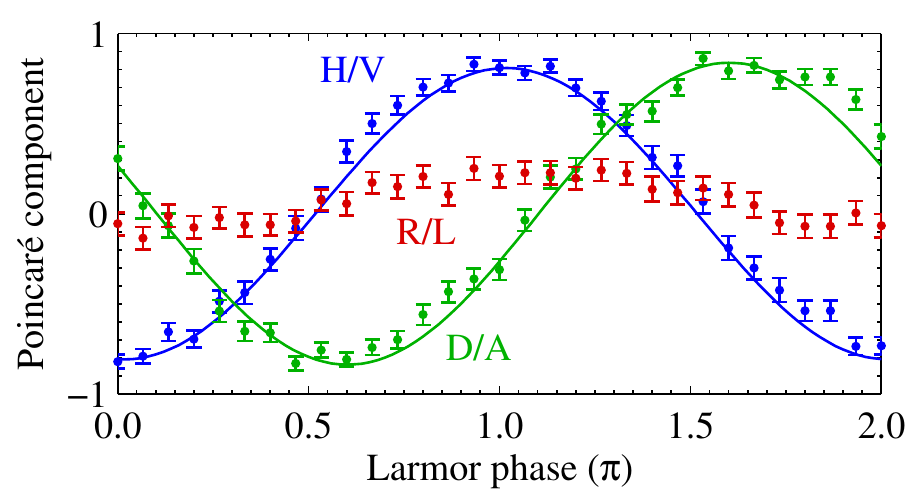} \label{fig:poincare}}
   \caption{Polarization analysis of single 393\,nm photons. (a) Conditional photon arrival-time histograms, shown for photons projected onto the horizontal (blue) and vertical (green) polarization state, and conditioned on the projection of the ion onto the state $\ket{+}$. The red curve shows the unconditioned arrival-time distribution, i.e.\ the sum of the blue and green histograms. The shaded area represents the time window used for data analysis. The bin size is 5\,ns and the overall measuring time is 20\,min. (b) State tomo\-graphy of the photonic polarization for the same atomic-state preparation and projection as in (a). Shown are the three Poincar\'e components as functions of the Larmor phase. The solid lines are sinusoidal fits.}
\end{figure}

A representative data set is shown in Fig.\,\ref{fig:oscphotons}. In this case, the atomic state is projected onto the superposition $\ket{+}=\frac{1}{\sqrt{2}}\left(\ket{-\half}+\ket{+\half}\right)$, and horizontally (H) and vertically (V) polarized photons are detected. The observed oscillations in the arrival-time distribution $P(t)$ agree, up to a reduced visibility, with the behaviour expected from the model \cite{Schug2014}, 
\begin{equation}
   P_\text{V,H}(t) \propto 1 \pm \cos(\varphi_\text{729}+\omega_\text{L}t)~.
   \label{eqn:arrivaltime}
\end{equation}
The oscillation period of about 64\,ns corresponds to the Larmor precession with $T=\frac{h}{4\,\mu_B B}$ expected for our static magnetic field $B=2.8\,\text{G}$ and taking into account the Land\'e factors of the $\sstate$ and $\Dstate$ manifolds. The overall envelope of $P(t)$ displays the exponential decay (with the Raman scattering rate) of the probability for the atom to remain in $\Dstate$ while it is exposed to the 854\,nm photons. 

From data as in Fig.\,3(a), we perform full tomography of the photonic polarization. We project the photon onto the three bases of linear (H/V and D/A) and circular (R/L) polarization by appropriate settings of the half-wave and quarter-wave plate. As a trade-off between high fidelity of the photonic state and high detection efficiency, we select detection events within the first 450\,ns of the arrival-time distribution, and we display them as a function of the Larmor phase $\varphi_\text{L} = 2\pi (\frac{t}{T}\,\text{mod}\,1)$. Fig.\,\ref{fig:poincare} shows the Poincar\'e components of the reconstructed polarization state. As expected, the linear polarization rotates with the Larmor phase, and the two linear bases show a $\frac{\pi}{2}$ phase difference. The ellipticity of the polarization (R/L component) is close to zero, with a slightly visible oscillation originating from a non-perfect calibration of the wave plates.

\subsection*{Entangler operation}

\begin{figure}[ht]
   \includegraphics[width=0.6\columnwidth]{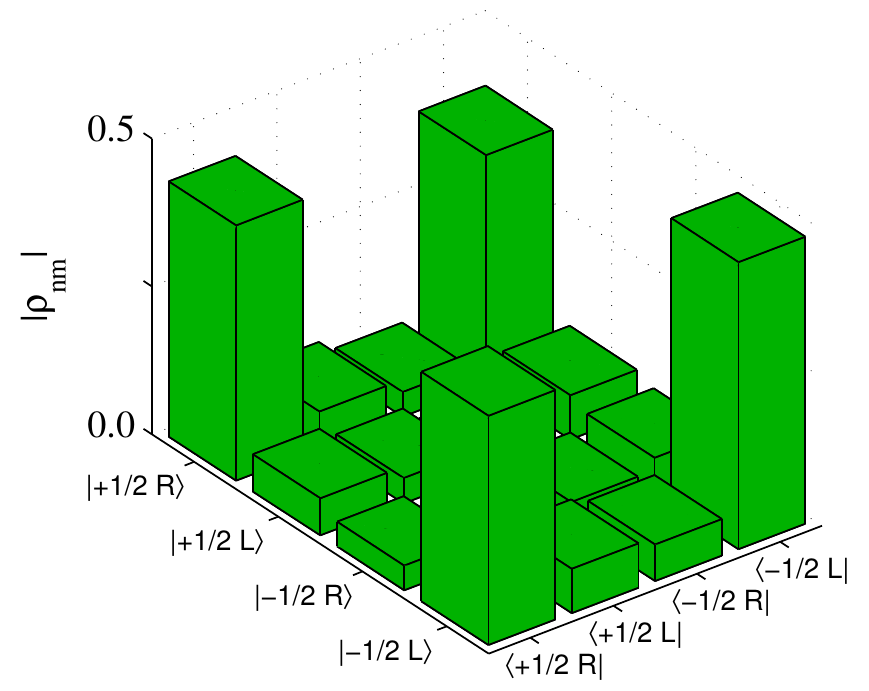}
   \caption{Density matrix of the entangled ion-photon state in the product basis. Shown are the moduli of the matrix entries.}
   \label{fig:densitymatrix}
\end{figure}

In order to verify the operation of our interface as a source of entangled ion-photon states, we perform standard quantum-state tomography \cite{James2001} of the two-qubit system through correlation measurements in the product bases of photonic and atomic qubit. Using a maximum-likelihood approach \cite{James2001}, we reconstruct the physical quantum state that is most likely to have produced the experimental data. From the derived density matrix, depicted in Fig.\,\ref{fig:densitymatrix}, we find a fidelity $F=84.6(2)\,\%$ with respect to the maximally entangled state (Eq.\,(\ref{eqn:entangledstate})) for our 450\,ns time window. This value exceeds the classical threshold of $\frac{2}{3}$ by more than 80 standard deviations and thus clearly indicates the creation of entanglement at the output of our interface.

The two main sources of infidelity in our set-up are currently magnetic-field fluctuations which limit the atomic coherence time and, as mentioned above, non-perfectly calibrated wave plates. The photon-detection efficiency for this experiment amounts to 0.353(1)\,\%, including that only 50\,\% of the arriving photons are transmitted through the polarization analyzer. At 11\,kHz repetition rate and taking the PMT quantum efficiency of 28(1)\,\% into account, we obtain $140(5)\,\text{s}^{-1}$ fiber-coupled photons that are entangled with the ion; removing the polarization analysis would double that rate.

\subsection*{Sender operation}

Operation of our interface for ion-to-photon quantum-state transfer is accomplished when a prepared atomic input state, characterized by angles $\vartheta_\text{D}$ and $\varphi_\text{D}$ (Eq.\,\ref{eqn:initialstateD}) leads to the emission of a 393\,nm output photon in the corresponding polarization state. This interface operation involves projection of the final atomic state onto a fixed superposition, for which we choose the $\ket{+}$ state. In this case the 393\,nm photon detected at time $t$ will be described by 
\begin{equation}
   \ket{\psi_\text{393}} = \cos\frac{\theta}{2} \ket{\text{R}} - \sin\frac{\theta}{2} e^{i\phi} \ket{\text{L}}
\end{equation}
with $\theta=\vartheta_\text{D}$ and $\phi=\varphi_{729}+\omega_\text{L} t$. One sees that the Larmor precession of the input state enters into the polarization of the output state. This is convenient for verification of the interface (see below); it has to be taken into account as a time-dependent phase shift when the interface is operated in a quantum communication scenario. 

\begin{figure}[ht]
   \makeatletter\renewcommand{\@thesubfigure}{\hspace{4mm}\smash{\thesubfigure}}\makeatother
   \subfigure[]{\includegraphics[width=0.35\columnwidth]{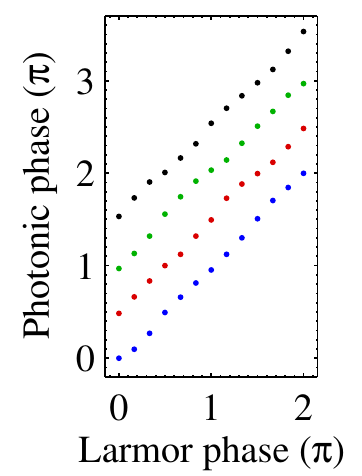} \label{fig:phaselines}}
   \hspace{-4mm}
   \makeatletter\renewcommand{\@thesubfigure}{\hspace{8mm}\smash{\thesubfigure}}\makeatother
   \subfigure[]{\raisebox{2mm}{\includegraphics[width=0.55\columnwidth]{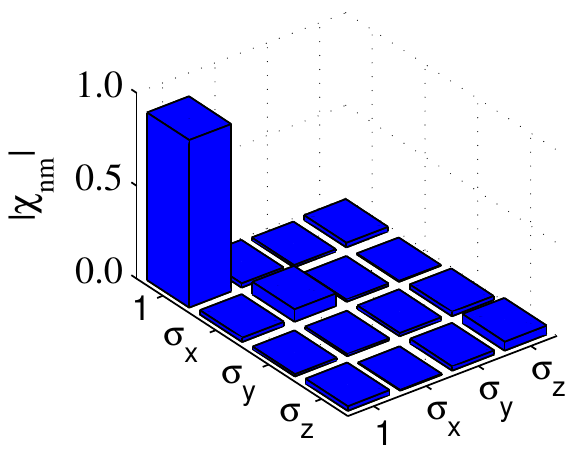}} \label{fig:qpmatrix}}
   \makeatletter\renewcommand{\@thesubfigure}{\hspace{5mm}\smash{\thesubfigure}}\makeatother
   \subfigure[]{{\includegraphics[width=0.9\columnwidth]{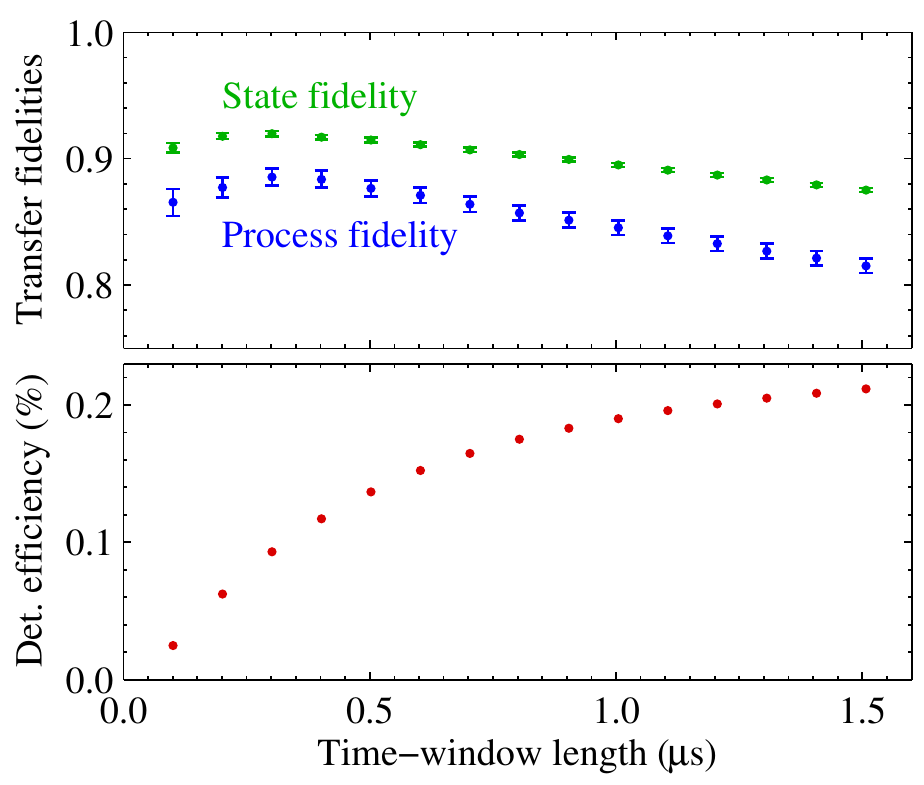}} \label{fig:fidelities}}
   \caption{Characterization of ion-to-photon quantum-state transfer. (a) Phase of the photon's linear polarization as a function of the Larmor phase and for $\varphi_{729}=\{0,\frac{\pi}{2},\pi,\frac{3\pi}{2}\}$ (top to bottom). (b) Quantum-process matrix in the Pauli basis. Shown are the moduli of the matrix entries. (c) Process fidelity (blue dots), mean quantum-state fidelity (green dots) and detection efficiency for different lengths of the detection time window. Where not indicated, error bars are smaller than the size of the symbols.}
\end{figure}

First, we prepare the four symmetric ($\vartheta_\text{D}=\frac{\pi}{2}$) superposition states with $\varphi_{729}=\{0,\frac{\pi}{2},\pi,\frac{3\pi}{2}\}$ and analyze the photonic phase, i.e., the direction of the linear polarization of the 393\,nm photon. The result is displayed in Fig.\,\ref{fig:phaselines}: the linear dependence with steps of $\frac{\pi}{2}$ between the four lines shows that the photonic phase reflects faithfully both the Larmor phase $\omega_\text{L} t$ and $\varphi_{729}$.

In a second step, we perform full quantum-process tomography of the atom-to-photon state mapping process by using also the bare input states $\ket{\pm\fivehalf}$ and analyzing the photonic state in all three polarization bases. This allows us to reconstruct the process matrix $\chi$, defined through $\rho_\text{out}=\sum_{m,n}\chi_{mn}\sigma_m\,\rho_\text{in}\,\sigma_n$ with the Pauli matrices $\left\{\sigma_{i=1,..,4}\right\}=\left\{\mathbbm{1},\sigma_x,\sigma_y,\sigma_z\right\}$ \cite{Chuang1997}. The matrix is displayed in Fig.\,\ref{fig:qpmatrix}. The value $\chi_{11}$ represents the identity part of the quantum process and is known as the process fidelity. From our experimental data, we derive $\chi_{11}=90.2(1.0)\,\%$. Another measure for the quality of the state transfer is the mean fidelity of the reconstructed photonic states with respect to the atomic input states; averaged over all six input states, it amounts to 92.4(3)\,\%. It is instructive to investigate how the two figures of merit behave as a function of the detection time window, which is depicted in Fig.\,\ref{fig:fidelities}; the same figure shows the trade-off between fidelity and success probability of the interface. For the 450\,ns time window, we achieve 0.127(1)\,\% mean photon detection efficiency. At 10\,kHz repetition rate, we obtain $45(2)\,\text{s}^{-1}$ fiber-coupled photons. The efficiency reduction compared to the previous measurement is caused by using different wave plates that are better calibrated but have smaller aperture. Removing the wave plates and polarizer would allow us to transfer the atomic qubit state onto fiber-coupled photons at an estimated rate of $\sim 300\,\text{s}^{-1}$, similar to that for ion-photon entanglement.

An important source of infidelity for both ion-photon entanglement and ion-to-photon quantum-state transfer are magnetic-field fluctuations in the vicinity of the trap apparatus that limit the coherence of the atomic qubit state. We expect to mitigate this issue by means of an active magnetic-field stabilization. Additional minor sources of infidelity are detector dark counts ($\sim 30\,\text{s}^{-1}$) and spontaneous decay from the $\Pstate$ state back to $\Dstate$ (with 5.87\,\% probability \cite{Gerritsma2008}).

\subsection*{Conclusions}

In conclusion, we demonstrated a single-atom to single-photon quantum interface that is programmable to serve for bi-directional quantum-state transfer and generation of atom-photon entanglement, essential operations in quantum networks, especially for a quantum repeater. Entangled ion-photon pairs have been created at close to 300\,s$^{-1}$ rate (entangler operation). The reconstructed two-qubit state exhibits 84.6(2)\,\% fidelity with respect to a maximally entangled state. Using the same ingredients, we realized ion-to-photon quantum-state transfer (sender operation): an arbitrary atomic qubit state is mapped onto the polarization state of the output photon with 90.2(1.0)\,\% quantum-process fidelity and 92.4(3)\,\% mean quantum-state fidelity. Application of the same interface for heralded photon-to-atom quantum-state transfer (receiver operation) had been shown earlier \cite{Kurz2014}. Finally, the converter mode of operation of the interface provides single-photon frequency conversion, with a single atom replacing the conventional non-linear optical device \cite{Ikuta2011, Zaske2012}. Our experimental implementation is based on a Raman transition in a single $\Ca$ ion and is readily adapted to other ions and neutral atoms such as Rb. Its versatility and simplicity makes the interface a valuable tool for quantum repeater and hybrid (see, e.g., \cite{Meyer2015}) quantum networking technology.


\vspace{0.3cm}
\noindent \textbf{Acknowledgments.} We gratefully acknowledge partial support by the BMBF (QuOReP project, QSCALE Chist-ERA project, Q.Com-Q project).

\bibliography{bibliography}

\end{document}